1

# Variants of Lüscher's fermion algorithm

Artan Boriçi[*] and Philippe de Forcrand[a]

[a]Interdisciplinary Project Center for Supercomputing,
Swiss Federal Institute of Technology Zürich,
IPS, ETH-Zentrum, CH-8092 Zürich

We study improved variants of Lüscher's algorithm, including the cases of staggered fermions and finite density.

## 1. INTRODUCTION

Simulation of full QCD, though essential, remains very expensive because of the large number of operations required by standard algorithms like Hybrid Monte Carlo (HMC) [1]. Cheaper algorithms are needed. Their cost should scale better than that of HMC as the volume of the system grows and the quark mass decreases. The algorithm proposed by Lüscher [2], which purports to be a controlled approximation to full QCD, represents a viable alternative to HMC [3,4]. However it can be simply modified into more efficient, exact variants [5].

We summarize here our work on these variants, and extend the formulation to staggered fermions and finite density QCD.

## 2. LÜSCHER'S METHOD

The method [2,3] approximates the fermionic determinant of 2 degenerate quark flavors as

$$det Q^2 \approx \frac{1}{det P(Q^2)} \quad (1)$$

where $P(x)$ is a polynomial of even degree $n$ approximating $1/x$ in the interval $(0,1]$. Knowing the polynomial roots $z_k, k = 1,..,n$ one can write

$$det P(Q^2) = const \times \prod_{k=1}^{n/2} det(Q^2 - z_k)^\dagger (Q^2 - z_k) \quad (2)$$

The inverse of each determinant on the r.h.s. can be expressed as a Gaussian integral over a bosonic field. We denote the Dirac matrix by $D = (\mathbf{1} - \kappa M)$, where $M$ is the lattice Wilson hopping operator and $\kappa$ the hopping parameter. Lüscher chooses for $Q$ the hermitian operator $Q = c_0 \tilde{Q}$, where $\tilde{Q} = \gamma_5 D$ and $c_0 = (1 + 8\kappa)^{-1}$. $P(x)$ is constructed from Chebyshev polynomials and depends on an adjustable parameter $\varepsilon \in (0, 1]$. Its error is bounded uniformly in the segment $[\varepsilon, 1]$ [3]:

$$|1 - xP(x)| \leq 2 \left(\frac{1 - \sqrt{\varepsilon}}{1 + \sqrt{\varepsilon}}\right)^{n+1} \quad (3)$$

We assess the error of this method by measuring the fluctuations of $det\, Q^2 P(Q^2)$ around 1, by calculating the eigenvalues of $Q^2$ over a sample of $4^4$ configurations [5].

## 3. EVEN-ODD SPLITTING

By even-odd splitting the lattice sites, we factorize $det Q^2$ into equal, even and odd factors [5]:

$$det Q^2 = det\, c_0^2 (\mathbf{1} - \kappa^2 M^2)^2_{even} \quad (4)$$

with $c_0 = (1 + 64\kappa^2)^{-1}$. It saves a factor 2 in memory, but a bosonic update requires about as many operations as before. Nonetheless the smallest eigenvalue increases as seen from Fig.1. In Fig.2 we show the error decreasing faster with $n$ after even-odd splitting. The gain is a factor 2 to 3.

One can achieve further gains by rescaling $Q$ to $Q/c_M$. In [3] is assumed $c_M \geq 1$ to guarantee that the spectrum is bounded by one. But for $\beta$ finite one can take $c_M < 1$, and the smallest eigenvalue of $Q^2$ will increase, leading to better convergence. Thus in Fig.2 we tuned $c_M$ to 0.6 instead of its default value of 1.

[*]This work is done under Grant No. 2100 - 037744.93 of the Swiss National Foundation for Scientific Research

Figure 1. Left part: a typical spectrum of the Dirac matrix $D$ in the complex plane ($4^4$ lattice, $\beta = 6$, $\kappa = 0.14$). Right part: the spectrum is shown after even-odd preconditioning.

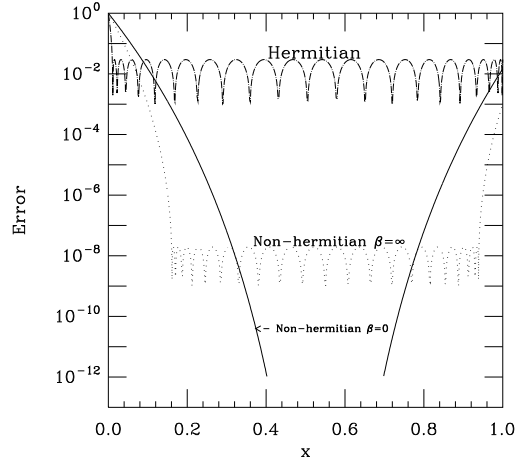

Figure 3. The magnitude of the error of the polynomial approximation $P(x)$ as $x$ varies from 0 to 1. The three approximations shown are the original hermitian approximation of Lüscher, the non-hermitian approximation inside a circle as appropriate for $\beta = 0$, and the non-hermitian approximation inside an ellipse of aspect ratio 2, as appropriate for $\beta = \infty$. All cases correspond to the same quark mass.

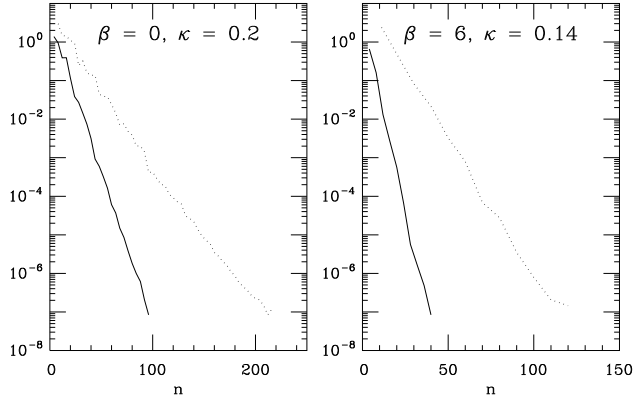

Figure 2. The magnitude of the error on the determinant as a function of $n$, in the original formulation (dotted line) and after even-odd preconditioning (solid line).

## 4. NON-HERMITIAN VARIANT

We look for an approximation $detD \approx 1/detP(D)$ where now $P(z)$ of degree even is defined in the complex plane. Applying (2) for $D$ and using the fact that $D = \gamma_5 D^\dagger \gamma_5$ and $\gamma_5^2 = \mathbf{1}$ one gets

$$detP(D) = const \times \prod_{k=1}^{n/2} det(D - z_k)^\dagger (D - z_k) \quad (5)$$

This approximation breaks down for $detD$ negative, which can only occur for very small quark masses. By this non-hermitian formulation one can simulate an odd number of quark flavors (or any number of non-degenerate flavors).

If the complex spectrum of $D$ is bounded by the ellipse centered at $(d, 0)$, with large semiaxis $a$, and focal distance $c$, and if we construct $P(z)$ by Chebyshev polynomials, we get [5]

$$|1 - zP(z)| \leq 2 \left( \frac{a + \sqrt{a^2 - c^2}}{d + \sqrt{d^2 - c^2}} \right)^{n+1} \quad (6)$$

for all $z$ within the ellipse defined above. We show in Fig.3 the magnitude of the error along the real axis for $n = 20$ and the quark mass $= 0.1$. In the non-hermitian case the error falls off exponentially until it becomes uniformly oscillating in some interval. This is in sharp contrast with the hermitian approximation of Lüscher.

We measure the fluctuations of $(DP(D))^2$ by calculating the eigenvalues of $D$ applying the same methodology as before. In Fig.4 we show that the gain over the hermitian case is up to a factor 4 to 5 at $\beta = 0$. Note that simulating one flavor to the same accuracy would require half as many bosonic fields $n$.

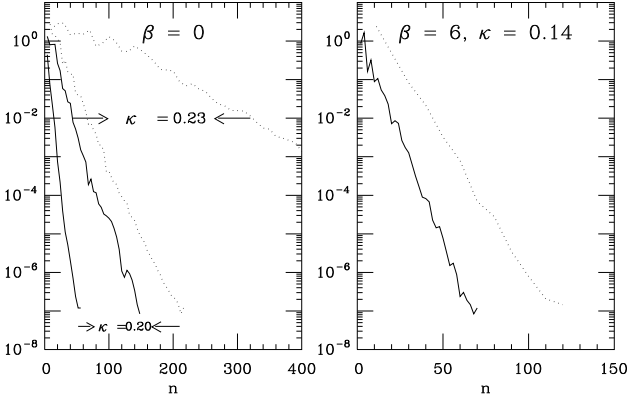

Figure 4. The magnitude of the error on the determinant as a function of $n$, in the original formulation (dotted line) and in the non-hermitian variant (solid line).

## 5. METROPOLIS TEST

Lüscher's original proposal includes the monitoring of the error, and the possibility of obtaining exact results by re-weighting the Monte Carlo measurements of each observable. It is more attractive however to correct for the error through a Metropolis test, just like HMC corrects for the error of the guiding Hamiltonian.

A first attempt along these lines [6,4] used the Lanczos algorithm to obtain eigenvalues of $Q^2$. The error $det\ Q^2 P(Q^2)$ is then easy to calculate. However the cost of this procedure grows prohibitively like the square of the volume $V$ of the lattice, unless one accepts an "error on the error".

We proposed instead [5] to use a noisy, unbiased estimator of the ratio of errors between the new and the old configurations. The ratio to estimate is $det(D'P(D'))^2 / det(DP(D))^2$, calling $D'$ and $D$ the Dirac operators for the new and old configurations respectively. Taking the denominator as a partition function, this ratio can be rewritten

$$< e^{-\eta^\dagger (W^\dagger W - \mathbf{1})\eta} > \qquad (7)$$

where the average $<>$ is taken over all Gaussian vectors $\eta$, and $W = [D'P(D')]^{-1} DP(D)$. It is sufficient to estimate this ratio by taking one Gaussian $\eta$ only at the end of each "trajectory", thus solving one linear system as in HMC. It is easy to show that the associated cost, expressed in number of update sweeps, is constant as the volume of the system grows or the quark mass decreases. The proof of detailed balance is more subtle.

Consider a generic update $U \to U'$ of the gauge fields, consisting of a "trajectory" of Lüscher updates followed by a Metropolis step. Detailed balance is satisfied if

$$\frac{P_L(U \to U')}{P_L(U' \to U)} \ \frac{P_{acc}(U'|U)}{P_{acc}(U|U')} = \frac{e^{-S_g(U')} det^2(D')}{e^{-S_g(U)} det^2(D)} \quad (8)$$

where $P_L$ and $P_{acc}$ are the probabilities for the set of Lüscher steps and the Metropolis step. Let us assume that the Lüscher updates satisfy detailed balance with respect to the Lüscher action $e^{-S_g} det^{-2}(P(D))$. This can be simply achieved by taking care that the order of the successive $U$ and $\phi_k$ updates in a trajectory can be reversed. Then eq.(8) implies

$$\frac{P_{acc}(U'|U)}{P_{acc}(U|U')} = \frac{det^2(D'P(D'))}{det^2(DP(D))} \qquad (9)$$

If one choses one Gaussian vector $\eta$ and simply takes $P_{acc}(U'|U) = min(1, e^{-\eta^\dagger (W^\dagger W - \mathbf{1})\eta})$, then $P_{acc}(U|U') = min(1, e^{-\eta^\dagger ((W^{-1\dagger} W^{-1}) - \mathbf{1})\eta})$, and detailed balance will *not* be satisfied.

However it is simple to restore detailed balance with the following prescription:
- if, say, $S_g(U') > S_g(U)$,
$P_{acc}(U'|U) = min(1, e^{-\eta^\dagger (W^\dagger W - \mathbf{1})\eta})$
- otherwise,
$P_{acc}(U'|U) = min(1, e^{+\eta^\dagger ((W^{-1\dagger} W^{-1}) - \mathbf{1})\eta})$

## 6. STAGGERED FERMIONS

Lüscher's method can be applied to staggered fermions. In this case the Dirac matrix has the form $D = m\mathbf{1} + iB$, where $m$ is the bare mass and $B$ is a hermitian matrix with extreme eigenvalues $\pm \lambda$. Then, in the hermitian formulation we have $Q^2 = D^\dagger D = m^2 \mathbf{1} + B^2$. Using (6) with $a = c = \lambda^2/2$ and $d = m^2 + \lambda^2/2$ the error bound will be

$$|1 - zP(z)| \leq 2 \left( \frac{\lambda}{m^2 + \sqrt{m^2 + \lambda^2}} \right)^{2(n+1)} \qquad (10)$$

for $z \in [m^2, m^2 + \lambda^2]$.

Similary as for Wilson fermions, one can also define a non-hermitian approximation using (2) for $D$ and the identities $D = \Sigma D^\dagger \Sigma$, $\Sigma^2 = \mathbf{1}$, where $\Sigma$ is $\pm 1$ on even/odd sites. In this case, taking $a = 0$, $c = i\lambda$, and $d = m$ we get the same error bound as above, but now for $z \in [m - i\lambda, m + i\lambda]$.

Thus, there are no savings from a non-hermitian approximation. The situation is the same as for the calculation of staggered fermion propagators, where no savings over CG can be achieved by a non-hermitian solver [7].

## 7. FINITE DENSITY QCD

There are several difficulties to simulate finite density QCD: i) since the determinant is not real anymore, its representation as a Gaussian integral over bosonic fields may not be convergent. Additional problems are: ii) how to get the phase of the determinant without calculating it directly and iii) how to deal with the large MC fluctuations caused by the complex phase (sign problem). We address here only problems i) and ii).

If we express $exp(\mu) = cosh(\mu) + sinh(\mu)$, then we can decompose the quark matrix as $D = D_1 + D_2$, where $D_2$ contains the $sinh(\mu)$ term and $D_1^\dagger = \gamma_5 D_1 \gamma_5$, $D_2^\dagger = -\gamma_5 D_2 \gamma_5$. If we assume an even degree polynomial $P(z)$ defined in the complex plane and apply (2) for $D_1 + D_2$, we get

$detP(D) =$

$const \times \prod_{k=1}^{n/2} det(D_1 - D_2 - z_k)^\dagger (D_1 + D_2 - z_k) =$

$const \times \prod_{k=1}^{n/2} det[(D_1 - z_k)^\dagger (D_1 - z_k) - D_2^\dagger D_2 +$

$(D_1 - z_k)^\dagger D_2 - D_2^\dagger (D_1 - z_k)] \qquad (11)$

In the last expression $(D_1 - z_k)^\dagger (D_1 - z_k) - D_2^\dagger D_2$ is hermitian, whereas $(D_1 - z_k)^\dagger D_2 - D_2^\dagger (D_1 - z_k)$ is antihermitian. There exists an interval of $\mu$ around 0 for which the hermitian part is positive and thus the corresponding bosonic Gaussian integral is convergent. In this case we can estimate the magnitude of the determinant from the hermitian part of (11), whereas the antihermitian part will give an estimation of its phase. This way we can approximate the determinant without calculating it directly.